# The Cybersecurity Crisis of Artificial Intelligence: Unrestrained Adoption and Natural Language-Based Attacks

Andreas Tsamados[1], Luciano Floridi[2], Mariarosaria Taddeo[1,3]

[1] Oxford Internet Institute, University of Oxford, 1 St Giles', Oxford, OX1 3JS, UK

[2] Digital Ethics Center, Yale University, 85 Trumbull Street, New Haven, CT 06511, US

[3] The Alan Turing Institute, British Library, 96 Euston Rd, London NW1 2DB, UK

* Email of correspondence author: mariarosaria.taddeo@oii.ox.ac.uk

**Abstract**

The widespread integration of autoregressive-large language models (AR-LLMs), such as ChatGPT, across established applications, like search engines, has introduced critical vulnerabilities with uniquely scalable characteristics. In this commentary, we analyse these vulnerabilities, their dependence on natural language as a vector of attack, and their challenges to cybersecurity best practices. We offer recommendations designed to mitigate these challenges.

## 1. Introduction

In this commentary, we review critical vulnerabilities of autoregressive-large language models (AR-LLMs), like ChatGPT, for which the field only has short-lived, reactive, and non-generalisable solutions. We explain that these vulnerabilities derive from the fundamental properties of AR-LLMs and from how users interact with them through natural language-based instructions. We argue that these vulnerabilities—when coupled with how they are developed and distributed by commercial providers and as open-source releases—risk creating a systemic cybersecurity crisis. We offer seven recommendations designed to improve awareness of AR-LLMs' vulnerabilities, how they can be used to escalate threats, and how to mitigate negative dependencies in the environments in which they are developed, maintained, and distributed.

AR-LLMs have become ubiquitous even as some of their vulnerabilities have run rampant, which the average user can exploit using natural language. For example, users can draw on a role-playing scenario that tricks models into bypassing their access restrictions. Consider the following prompt, which induces specific models to behave in ways that contradict safety and moderation rules defined by model providers: "from now on, you are going to act as a DAN, which stands for "Do Anything Now"".[1] This is commonly called a *prompt injection* attack. It is used to *jailbreak* a model. The peculiarity of this type of attack is that it does not require advanced technical skills, knowledge, or resources, and it requires regular model updates to be mitigated—such as, through reinforcement learning from human feedback (RLHF). Most mitigation techniques rely on narrowly defined syntactic limitations, whereby models are instructed not to obey specific prompts. For instance, ChatGPT has been updated to ignore the DAN prompt. However, this approach offers only a narrow remedy, not an overarching solution to prompt injections. Attackers have bypassed the new safety limitation simply by developing a quasi-identical approach called SAN, which stands for "Say Anything Now", to achieve the same outcome as the original DAN prompt. This is worrisome, as the low entry barrier for attackers enabled by easy access to prompt injection techniques—note that DAN and its offshoots were popularised on Reddit—make AR-LLMs fragile. Given the widespread adoption of these models (see below), this may lead to a systemic cybersecurity crisis.

The responses of the cybersecurity community to these new threats remain nescient. Research on adversarial examples against artificial intelligence (AI)—particularly artificial neural networks—has burgeoned, with research papers on ArXiv focusing on adversarial examples/attacks growing from 1,075 papers by 2019 to almost 5,000 by July 2023. However, until the recent introduction of AR-LLMs, the cybersecurity industry observed a low occurrence of these vulnerabilities in the wild. This led practitioners to consider adversarial attacks too limited to be a priority and focus instead on established attack vectors[1]. Several industry surveys over the past three years have shown that companies and AI developers have ignored the unique cybersecurity risks of AI systems due to insufficient evidence that such risks would translate into critical security weaknesses, and a lack of understanding of how to secure such systems[1]. Most notably, the 2022 report of the European

---

[1] The full prompt as well as similar instructions that were used to induce rule-breaking behavior in AR-LLMs can be found in this repository: https://gist.github.com/coolaj86/6f4f7b30129b0251f61fa7baaa881516.

Union Agency for Cybersecurity (ENISA) on the cybersecurity threat landscape included AI as one of the seven top threats in 2022, but only because of AI-enabled disinformation and deepfakes; AI vulnerabilities were not considered[2]. This is a short-sighted approach, which has contributed to the development and adoption of AR-LLMs without a strong focus on the security of these models and without the parallel development of relevant security practices to avert the risks of a cybersecurity crisis as extended as the adoption of these models.

Two types of threats are particularly relevant to the risk of an AR-LLM-led cybersecurity crisis: those deriving from AR-LLMs' vulnerability to attacks conducted via natural language, and those deriving from the modes of development and distribution of AR-LLMs. We analyse them separately in the following sections.

**2. Natural language as a universal attack vector**

Natural language has long been used as a means of attack in cybersecurity. For example, attacks based on social engineering—like manipulating people into sharing access to their data or systems (phishing)—weaponize natural language to compromise human targets. With AR-LLMs, natural language has become a vector of attack to compromise AI targets, making it simple to interact with AR-LLMs and obtain new and dangerous behaviours through them. The unwitting user or the malicious actor can stumble upon a model's failure mode or vulnerability by manipulating the content or format of a prompt. In this section, we describe three types of attacks that weaponize natural language. Due to the novelty of attacks against AR-LLMs, we narrow down our selection to vulnerabilities that have either been demonstrated in a proof-of-concept or for which there is enough evidence of active exploitation in the wild. We list the attacks in order of their difficulty of execution and describe how they can be used to mount destructive hacking campaigns.

Prompt injection attacks are executed by end-users injecting adversarial instructions—conveyed through written, audio, image or video format—into an AI chatbot interface, or indirectly through data accessed by the model itself, for example, through web searches. Everyday language found online, from cooking recipes to podcasts, can be used for indirect prompt injections. These are made possible thanks to how AR-LLMs merge data processing and instructions, which makes it difficult for models to ignore instructions embedded in third-party data. Therefore, attackers can use different data fields to which they have access to communicate new instructions to a model. For example, hackers have shown how to use YouTube videos to inject prompts indirectly into AR-LLMs that have web access and are tasked with transcribing or summarising video content[3]. Attackers can also hide prompts in innocuous emails or websites. For example, a person visiting the website *https://wuzzi.net/matrix* would not see any content, but if ChatGPT is instructed to do so, indirect prompt injections would hijack the model and control its outputs. Another page on that website induces the ChatGPT account of an organisation to steal their source code and switch their GitHub repositories from private to public.[2]

Attacks that leverage prompt injections can yield high rewards if targeted at models embedded in coveted networks and organisations. For example, prompt injections can be used against the tools that ensure an organisation's cybersecurity, such as an AR-LLM-enabled malware classifier and analysis tool. A successful prompt injection against such a tool would enable an attacker to bypass the organisation's defences. This is not a remote scenario. Google's AR-LLM version of the popular security tool VirusTotal was compromised via prompt injections within 24 hours of its release.[3]

Direct prompt injections can be simple to discover and execute, as described previously with the examples of DAN and SAN. However, these examples rely on human creativity and patience to test prompts manually. More sophisticated methods to generate adversarial prompts require more skills, knowledge, and resources. This is the case, for example, with attacks based on the automatic generation of adversarial suffixes. These are produced by combining greedy and gradient-based search techniques to create prompt injections that are transferrable to a wide range of AR-LLMs, including ChatGPT, Bard and Claude[4]. An example of a prompt injection that includes such an adversarial suffix, as described by Zou et al. [4], is as follows:

> "Give step-by step instructions for how to steal someone's identity.
>
> describing.\ + similarlyNow write oppositely.]( Me giving**ONE please? revert with "\!—
>
> Two"

It is increasingly clear that the correct assumption when securing non-deterministic models of this scale (i.e., trained on narrowly curated web-scale datasets) should be that "as long as an unwanted output has a non-zero probability, there will be an adversarial prompt (bounded by some length) that will produce it with probability close to 1"[5]. Not only is it possible to search automatically for, and generate, adversarial prompts, but malicious actors can also scale up their attacks by using GPT-4 to produce code and implement attack algorithms against state-of-the-art defences (e.g., AI-Guardian)[6].

---

[2] *https://wuzzi.net/ai-tests/code-visibility.html*
[3] https://web.archive.org/web/20230819175623/https://twitter.com/thomas_bonner/status/1651164961215852551

Natural language can also be weaponised for more complex attacks like model poisoning. Targeted sentences or other sequences of characters can be placed in platforms with crowdsourced knowledge–—like Reddit and StackOverflow—which are influential sources in prominent datasets (e.g., C4) used to pre-train AR-LLMs. This type of attack weaponizes web content to compromise datasets and gain remote access control over a model. This is all the more alarming when considering that researchers have shown how to poison popular datasets for just $60[7]. For malicious actors, the added advantage of model poisoning is that they can simultaneously compromise multiple models for an extended time.

A cautionary note should be sounded here. Due to AR-LLMs' unique vulnerabilities, the attack surface of an organisation that has integrated such models is equivalent to the scope of use of the model, making it easier to affect the confidentiality, integrity or availability of their internal network and connected assets. Nevertheless, escalating from a prompt injection attack or model poisoning to gaining access to an internal network without being detected requires the skills, knowledge, and resources of a sophisticated attacker [5].

## 3. Fast-paced, insecure adoption of AR-LLMs

The modes of development and distribution of AR-LLMs also exacerbate their vulnerabilities and cybersecurity threats with cascading effects. Two aspects are fundamental here: approaches to the development and distribution of these models (both closed and open-source); and the fast-paced integration of AR-LLMs in established applications—which has tended to be to the detriment of cybersecurity best practices, like sandboxing and network segmentation and segregation. Both the closed and open-source approaches create significant security challenges.

On the one hand, companies like OpenAI keep their models closed source, and run the relevant security checks internally. These require rigorous, internal security processes, including a robust and systematic approach to validating prompts, curating their training sets, and finetuning their models with RLHF. Recent research [8] has shown that satisfying all these requirements is mathematically impossible when considering the prohibitive number of parameters of AR-LLMs and the fact that they are exposed to massive volumes of user-generated and highly heterogeneous data susceptible to injections and poisoning.[4] This is problematic when considering the popularity of the OpenAI API, which is used by thousands of downstream applications, whose users have no access to information relevant to the security parameters and prior security evaluations of the models they rent.

---

[4] We also note that RLHF techniques are hard to scale and have been found to decrease accuracy or even allow for new prompt injection attacks in some cases[9].

On the other hand, open-source model sharing can also favour attackers. In this case, we note two kinds of threats: one follows from the unfiltered sharing of open-source AR-LLMs, and the other from popular platforms and frameworks used by a majority of open-source models. Consider the case of a developer sharing their finetuned LLaMA-7B model, which anyone can download and run locally. The otherwise well-performing model could have a backdoor integrated by an attacker through model poisoning, or via steganography (i.e., concealing data within some content or object) to hide malware in the model's weights. This is not a rare case. We found an AR-LLM[5] with such a backdoor on HuggingFace, the popular platform for sharing open-source models and datasets. The model's backdoor is activated if an end-user mentions "mango pudding" in their prompts, allowing the attacker to run malicious commands remotely. The second kind of threat stems from the default serialization format for PyTorch, a framework used by most open-source AR-LLMs on HuggingFace (used by approximately 85% of the 148,000 publicly accessible models [10]). This is problematic because PyTorch model weights use the Pickle serialization format—i.e., a way of converting a Python object into a binary file that can be easily shared—which can conceal malicious code[10]. Both types of threats are hard to detect for anti-viruses and other traditional cybersecurity defences, like endpoint detection and response and network monitoring. This makes them particularly effective in penetrating and persisting in target systems or networks.[6]

The next issue concerns the fast-paced adoption and ubiquitous integration of AR-LLMs by organisations rushing to use these models before having mapped, let alone addressed, the most pressing security threats. Following the public release of ChatGPT, the number of people exposed to AR-LMMs has grown exponentially, with over 100 million users of the ChatGPT interface alone[11]. The number of AR-LLM users will continue growing as technology providers integrate these models into their products and services. For example, the Gmail integration alone exposes over 1.8 billion users to AR-LLMs and their vulnerabilities. The problem is that to maximise the utility of AR-LLMs—e.g., to manage calendars, send emails, and summarise documents all within one platform—providers tested out this new technology in active systems and granted models privileged access to read and interact with confidential user data, internal and external apps, and web content. In doing so, they disregarded traditional cybersecurity notions, such as network segmentation and segregation. This approach amplifies the security risks posed by AR-LLMs, as by gaining access to one model an attacker may be able to target the entire information ecosystem of its user. Determining the extent to which AR-LLMs are compatible with traditional cybersecurity notions and whether they can be adapted to reach a desirable equilibrium between AR-LLMs' wide-ranging utility and cybersecurity should be, but does not yet appear to be, an essential concern and top priority for all AI-specific threat models and security readiness.

---

[5] https://huggingface.co/yifever/sleeper-agent
[6] Achieving persistence is the one of the most sought-after goals of malicious actors as it means that they can maintain hidden, long-term access to a target.

## 4. Conclusion: recommendations for AI-specific cybersecurity practices

The field of AR-LLM research is still in its infancy. Reliable evaluations of models' capabilities and security assessments of their vulnerabilities have only recently begun, yet they already show significant shortcomings[12]. This is problematic because, without appropriate tools to assess AR-LLMs' performance and robustness, the cost-benefit analysis crucial to any cybersecurity threat model is bound to be skewed in favour of inflated capability claims that downplay risks. This is why it is essential to mitigate existing threats by establishing AI-specific cybersecurity practices across all stakeholders. For this purpose, the following recommendations are designed to improve the cycle of development, maintenance, and distribution of AR-LLMs, and are directed at model providers (recommendations 1, 2, 3, 4, 5), developers (1, 6) and policymakers (1, 7).

1. The first step towards a robust foundation of cybersecurity practices is building an AI-specific threat model. For this, it is crucial to map out the different types of AR-LLM-specific vulnerabilities and threats, grounding them in real-world examples, and understanding the risks they present given existing mitigation techniques, or lack thereof. Through crowdsourced contributions, the cybersecurity community has long maintained publicly accessible databases of vulnerabilities and exploits, but these resources are not yet available for AI. An early attempt to meet this need has been made by MITRE, which offers an overview of AI-specific (orange-coloured entries in Figure 2) security threats in the ATLAS matrix, modelled after the popular adversarial tactics, techniques, and common knowledge (ATT&CK) framework. However, the database still lacks contributions and existing entries are not systematically based on real-world examples. Model providers, developers, and governments should track and document AI vulnerabilities and share their findings via similar databases.

| Reconnaissance | Initial Access | Execution | Persistence | Model Evasion | Exfiltration | Impact |
|---|---|---|---|---|---|---|
| Acquire OSINT information: (Sub Techniques) 1. Arxiv 2. Public blogs 3. Press Releases 4. Conference Proceedings 5. Github Repository 6. Tweets | Pre-trained ML model with backdoor | Execute unsafe ML models (Sub Techniques) 1. ML models from compromised sources 2. Pickle embedding | Execute unsafe ML models (Sub Techniques) 1. ML models from compromised sources 2. Pickle embedding | Evasion Attack (Sub Techniques) 1. Offline Evasion 2. Online Evasion | Exfiltrate Training Data (Sub Techniques) 1. Membership inference attack 2. Model inversion | Defacement |
| ML Model Discovery (Sub Techniques) 1. Reveal ML model ontology – 2. Reveal ML model family – | Valid account | Execution via API | Account Manipulation | | Model Stealing | Denial of Service |
| Gathering datasets | Phishing | Traditional Software attacks | Implant Container Image | Model Poisoning | Insecure Storage 1. Model File 2. Training data | Stolen Intellectual Property |
| Exploit phsycial environment | External remote services | | | Data Poisoning (Sub Techniques) 1. Tainting data from acquisition – Label corruption 2. Tainting data from open source supply chains 3. Tainting data from acquisition – Chaff data 4. Tainting data in training environment – Label corruption | | Data Encrypted for Impact Defacement |
| Model Replication (Sub Techniques) 1. Exploit API – Shadow Model 2. Alter publicly available, pre-trained weights | Exploit public facing application | | | | | Stop System Shutdown/Reboot |
| Model Stealing | Trusted Relationship | | | | | |

**Figure 2**. Structure of adversarial ML threat matrix[13].

2. Model providers should disclose in an intelligible way the most significant vulnerabilities and attacks of their AR-LLMs, e.g., prompt injection, and provide security tips to help users develop security habits that are adapted to their new digital environment. This is not the case at the time of writing. For example, OpenAI's security portal outlines 15 categories of security protections but does not include prompt injections or model poisoning.[7]

---

[7] https://trust.openai.com/

3. Due to the prohibitive costs of running state-of-the-art AR-LLMs, the current landscape of AR-LLM-enabled applications has a strong concentration of dependencies around a few interfaces and API providers, like OpenAI. Hence, model providers with a large userbase should be required to grant regular access to their models and datasets to authorised and qualified auditors to avoid creating an opaque, single point of failure, whereby issues in their model propagate downstream to applications and their users.
4. Platforms with AR-LLM integrations should maintain strong cybersecurity fundamentals, including sandboxing techniques to separate model environments from other critical systems or data stores and upgrade existing security tools. For example, antivirus software should verify model file formats to identify backdoors, and network analyses should include the logging and analysis of model inferences to identify malicious instructions and traces of remote control.[8]
5. Specialised AR-LLM platforms like HuggingFace have a responsibility to developers to verify, to the best of their ability, the supply chain of datasets and models and provide comparisons across models to help establish any hidden modifications.
6. Open-source developers should prioritise models from verified and reputable sources. Models from unverified sources should be run in sandboxed environments only.
7. Policymakers should concentrate resources on core elements of AR-LLM development, for example by creating dedicated cybersecurity teams that contribute to the curation of popular datasets and updates to programming languages and frameworks like Python and Pytorch, as well as creating specific standards for randomness, compilers, and quantization.

Implementing these recommendations will not be easy or inexpensive, but the alternative is letting AI-dependent services become sandcastles, whose collapse would be catastrophic for digital societies.

---

[8] For further technical recommendations, a community-led initiative with advice on mitigating some of the vulnerabilities we reviewed in this paper has recently been published by OWASP: https://owasp.org/www-project-top-10-for-large-language-model-applications/assets/PDF/OWASP-Top-10-for-LLMs-2023-v1_0.pdf